# Neuroradiological features of traumatic encephalopathy syndrome using MRI and FDG-PET imaging: a case series in Australia.

Rowena Mobbs*[1,2,3], Fatima Nasrallah[3], Xuan Vinh To[3], John Magnussen[1], Jennifer Batchelor[4], Edward Hsiao[2,5], Mark Walterfang[6,7,8]

[1] Macquarie Medical School, Macquarie University, New South Wales, Australia

[2] Mater Hospital Sydney, New South Wales, Australia

[3] The Queensland Brain Institute, The University of Queensland, Australia

[4] School of Psychological Sciences, Macquarie University, New South Wales, Australia

[5] Department of Nuclear Medicine and PET, Royal North Shore Hospital, New South Wales, Australia

[6] Neuropsychiatry Centre, Royal Melbourne Hospital, Melbourne, Australia

[7] Department of Psychiatry, University of Melbourne, Melbourne, Australia

[8] Edith Cowan University, Perth, Australia

* Corresponding Author: Dr. Rowena Mobbs
MQ Health
2 Technology Place, Macquarie University
New South Wales 2109
Phone: +61 2 98123720
Email: rowena.mobbs@mqhealth.org.au

# Abstract

Objectives: This study examined whether currently existing clinical structural magnetic resonance imaging (MRI) and fluorodeoxyglucose positron emission tomography ($^{18}$FDG-PET) capabilities and board-certified radiologists' reports and interpretations can assist with traumatic encephalopathy syndrome (TES) diagnosis.

Design: retrospective case series.

Setting: this study assessed six patients with TES criteria recruited from the Sydney area, Australia

Main outcomes: patients' clinical history and clinical presentation along TES diagnostic criteria, board-certified radiologist reports of structural MRI and $^{18}$FDG-PET.

Results: one patient was classified as possible CTE, and the others were classified as probable CTE with significant RHI exposure history and a spectrum of cognitive deficits and other neuropsychiatric disturbances consistent with TES diagnostic criteria. Most common radiological features included atrophy of posterior superior parietal region and Evans Index > 0.25. FDG-PET's pattern of common regions of hypometabolism and differing hypometabolism across participants suggested that FDG-PET and structural MRI have the potential for stratifying different TES stages.

Conclusions: this study highlights the potential of a combined clinical and radiological approach using solely current capabilities to improve TES diagnosis and suggests a larger study. Such expanded investigation is crucial for advancing the ante-mortem diagnosis and management of CTE.

**Summary:**

Chronic traumatic encephalopathy (CTE) is a debilitating neurodegenerative disease that can only be confirmed post-mortem. Its *in vivo* probabilistic diagnostic counterpart, traumatic encephalopathy syndrome (TES), relies on clinical assessment but only offer probabilistic classification. This case series identified radiological features identifiable by currently existing MRI and FDG-PET capabilities and board-certified radiologists that can assist in TES diagnosis in combination with clinical assessments. The imaging methods can also offer a potential method for grading TES progression. The findings may catalyse further research into diagnostic and progression-grading tools and providing a framework for early detection and treatment for TES/CTE.

**Introduction**

Chronic traumatic encephalopathy (CTE) is a neurodegenerative condition that is predominantly associated with repetitive head impact (RHI) [1]. However, its definitive diagnosis relies on post-mortem evaluation [2]. Since definite CTE conclusion is only possible post-mortem, CTE classification has limited uses in clinical settings with still-living patients with symptoms management and treatment needs. To address this gap, diagnostic criteria for traumatic encephalopathy syndrome (TES) was developed, which use a set of clinical characteristics that were retroactively identified from confirmed CTE patients to classify living patients along a spectrum of certainty for CTE pathology: from suggestive of, possible, and probable CTE [3]; the spectrum ranges from more sensitive but less specific to less sensitive but more specific towards identifying potential CTE patients. Regardless of CTE probability identified by the TES diagnostic criteria, the condition remains under-researched with no standard management plan for current patients [2].

Structural magnetic resonance imaging (MRI) continues to be a primary tool for detecting atrophy in prevalent neurodegenerative diseases such as Alzheimer's disease (AD) and frontotemporal dementia (FTD) [4]; there are advantages for use of structural MRI in clinical diagnosis, given that the only FDA-approved MRI method is the structural MRI-based NeuroQuant [5–7]. Several studies have applied pre-clinical structural MRI to TES participants (for reviews, see [8,9]). The findings ranged from increased ventricle-to-brain ratio (VBR) [10], higher Evans index (marker of ventricular volume) [11], increased ventricular volumes [12,13], and increased cavum septum pellucidum (CSP) prevalence, length, or ratio [12,14–18]. Grey matter (GM) related findings in TES included: atrophy/decreased volumes of subcortical grey matter [12], total grey matter [12], limbic system structures (i.e. amygdala [19], hippocampus

[12,14,19–21], and cingulate gyrus [19]), the frontal cortices [14], insula [14], orbital-frontal lobes [13], dorsolateral-frontal lobes [13], superior-frontal lobes [13], medial-temporal lobes [13,22], anterior-temporal lobes [13,14,21,22], and thalamus [16,23]. No significant morphological difference found was also observed in a couple TES studies [24,25].

Given the nature of CTE as a tauopathy condition, a few studies have explored the use of Tau-Positron Emission Tomography (PET) imaging of in RHI patients. FDDNP (2-(1-{6-[(2-[fluorine-18]fluoroethyl)(methyl)amino]-2-naphthyl}-ethylidene)malononitrile) were used in most early PET studies of RHI/CTE/TES [9]. However, FDDNP was reported to both bind to and insufficiently bond to amyloid plaques and neurofibrillary tangles *in vivo* [26]. In any case, FDDNP is not FDA-approved and there is limited evidence for its clinical value [9]. Flortaucipir ($^{18}$F) (FTP) is a FDA-approved radiotracer developed to detect neurofibrillary tangles characteristic of Alzheimer's disease (AD) [27]. It has been shown that FTP weakly binds to brain tissues with dense CTE pathology while more strongly binds to tissues with AD tau [28], indicating that while FTP may not be useful to detect CTE *per se*, it may be useful for ruling out AD as the causal disorder of TES symptoms. FTP also showed "off-target" and non-tau-related bindings to regions commonly involved in CTE [9], limiting its usefulness in Tau-PET for CTE/TES detection. On the other hand, 2-deoxy-2-($^{18}$F)fluoro-deoxyglucose (FDG)-PET has been a staple in the diagnostic workup for autopsy-confirmed dementia [29]. FDG-PET provides *in vivo* measurements of the degree of and spatial patterns of change in brain metabolism and presumably change in synaptic activity [9]. TES patients have been shown to have lower FDG in fronto-temporal and some parietal (precuneus and angular/supramarginal) areas compared to controls [14]. Small studies comparing former athletes with a history of RHIs but without clear TES diagnosis against controls showed lower brain metabolism in the posterior cingulate cortex [30], the parieto-occipital lobes [30], and frontal lobes [24,30]. In a couple of small [31] and qualitative study [14] of retired athletes with RHIs but without TES diagnosis, qualitative evaluation of FDG-PET showed hypometabolism in the medial and lateral temporal lobes, frontal lobes, and temporoparietal regions [31] or fronto-temporal areas [14].

In this case series, we describe and characterise six patients with significant exposures to RHI and meeting TES criteria and recruited from the North Sydney area, Australia. Their characterisation will be along the TES diagnostic criteria, clinically available imaging, treatments, and improvements.

**Materials and Methods**

Six TES patients (n = 6, 5 males and 1 female, age range 30–57) are included in this case series. All patients were presented to the Mater Hospital North Sydney neurology clinic (through referral by their primary healthcare provider) and were patients of author R.M. Patients' basic demography and RHI exposure information are detailed in Table 1. The patients underwent neurological examination, psychological and cognitive assessments, and MRI and PET imaging at the Mater Hospital. Patients with detectable cognitive decline (Table 2) and neuropsychiatric issues (Table 3) were provided standard neurological care and management (summarised in Table 3). Qualitative or semi-quantitative responses to medication and environmental measures were recorded using clinically available assessment tools, including the Addenbrooke's Cognitive Examination (ACE-III) and the Depression Anxiety Stress Scales 21 (DASS-21).

MRI scans were performed on a MR750W scanner (GE Medical Systems, Milwaukee, USA) with a 64-channel head coil at I-MED Radiology at the Mater Hospital North Sydney. High-resolution anatomical scans were conducted with a T1 3D

Gradient-Echo (GRE) BRAin VOlume (BRAVO) sequence with the following parameters: Repetition Time (TR)/Echo Time (TE) = 2/6 ms, matrix size = 512 × 512 × 194, and voxel resolution = 0.5 × 0.5 × 1 mm. T2-weighted dark fluid structural images were acquired using a sagittal 3D fast spin echo (CUBE) FLuid Attenuated Inversion Recovery (FLAIR) with the following parameters: TR/TE = 140/5500 ms, matrix size = 512 × 512 × 174, and voxel resolution = 0.5 × 0.5 × 1 mm. PET imaging was performed on a Discovery 690 CT/PET scanner (GE Medical Systems, Milwaukee, USA) with 2.0 MBq/kg $^{18}$F-FDG injection and 55 minutes uptake period. Blood glucose levels were measured to ensure the absence of hyperglycaemia during the FDG-PET scan. MRI and PET scans were interpreted and assessed by a panel of two radiologists specialising in neuroradiology and for the purpose of this study patients' imaging outcomes were based on the radiologists' reports.

Patients' clinical data, health records, and imaging reports are stored in the Clinic to Cloud practice management system (https://www.clinictocloud.com/). This case study uses a retrospective design and study ethics approval was obtained from the Macquarie University Human Research ethics committee (ethics approval number 6057) for data to be extracted from the database for analysis and publication conditioned upon patients' or substitute decision makers' informed consent for their deidentified data to be used. Patients were selected according to the NINDS-CTE diagnostic criteria [3] and the Major Neurocognitive Disorder in Diagnostic and Statistical Manual of Mental Disorders, fifth edition (DSM-V) [32]. Written informed consent was obtained from all patients; after informed consent was granted, patients' data and reports were extracted from the database for the study.

**Results**

Summary of all participants basic demography (Table 1), cognitive decline (Table 2),

and psychiatric issues (Table 3), structural MRI features (Table 4), and FDG-PET features (Table 5) are provided. Detailed description for each case along the TES criteria, imaging reports, treatments and improvements are as followed:

*Case 1:*

*TES criteria I:* A male in his 50s with a history of Traumatic ExposuRe (TER) of 33 years in contact sports, including martial arts, rugby, and boxing. 50 total recalled concussions (TRCs), 10 involving loss of consciousness, with the first at age 4.

*TES criteria II:*

  Cognitive impairment: Patient presented with a 5-year history of short-term memory impairment and mood disturbances. Initial ACE-III score was 85, with deficits in verbal fluency, processing, interpretation, recall, and insight. Neuropsychological testing showed severe episodic memory disorder, significant executive dysfunction, and impaired verbal abstract reasoning. Despite average to high average intelligence, the patient's verbal learning and recall were significantly below average, and the patient made numerous uncorrected errors.

  Neurobehavioral dysregulation: emotional lability, irritability, and anger outbursts, contrasting with his previously passive personality.

  Progressive course: patient experienced ongoing cognitive and neuropsychiatric decline in the three years following first presentation.

*TES criteria III:* the patients' symptoms were not fully accounted for by other disorders.

*TES criteria IV:* mild dementia.

*Other supportive features:*

Delay onset: recalled onset of symptoms started approximately 10 years after cessation of RHIs.

Motor signs: Neurological examination revealed mild bradykinesia, cogwheel, and axial rigidity, but no micrographia, frontal release signs, or cerebellar features.

Other psychiatric features: social isolation, anxiety, and depression. Alcohol Use Disorders Identification Test (AUDIT-C) [33] score of 4, and the patient showed signs of Rapid Eye Movement (REM) sleep behaviour disorder (RBD).

*Provisional level of certainty for CTE Pathology:* probable CTE.

*Neuroimaging findings:* MRI data revealed mild Posterior Superior Parietal Region (PSPR) atrophy and enlargement of the adjacent sulcus, suggesting grey matter volume loss in the PSPR. Ventricular widening, hippocampal, and precuneal volume loss were absent. PET findings indicated hypometabolism in the bilateral parietal, striatal, frontal, and precuneal regions, with the lateral temporal region having more overt hypometabolism than the medial temporal region; the maximal severity of hypometabolism across the whole brain was moderate.

*Treatments and improvements:* thiamine supplementation and melatonin for RBD. Patient responded well to Fluvoxamine selective serotonin reuptake inhibitors (SSRI) and Donepezil cholinesterase inhibitor and showed improvement in cognitive scores and mood.

### *Case 2:*

*TES criteria I:* A female in her 30s with a TER of 26 years of traumatic brain injury (TBI) due to weekly head banging and a history of sustained sexual victimization and physical violence, including strangulation and direct head trauma and over 100 TRCs.

*TES criteria II:*

Cognitive impairment: patient presented with a 5-year history of cognitive decline and reduced insight. Neurological examination revealed impairments in learning, performing complex tasks, attending, processing, and comprehension, including recognition of faces and poor motivation. ACE-III score of 87/100, with prominent difficulty in registration. Neuropsychological testing indicated high-average intellect but a significant disparity between verbal and nonverbal episodic memory. Verbal learning and retention were high average, while visual learning and retention were almost 2.0 standard deviations below. Patient also struggled with strategic problem solving, novel response generation, and processing speed.

Neurobehavioral dysregulation: verbal irritability.

Progressive course: patient experienced continued decline in cognitive ability.

*TES criteria III:*

Patient had diagnoses of major depressive disorder and post-traumatic stress disorder (PTSD) that were unlikely to explain the other clinical features and more likely to be the consequence of the history of TBIs, sexual victimization, and physical violence. Patient had no toxin exposure history.

*TES criteria IV:* subtle/mild functional limitation.

*Other supportive features:*

Delay onset: not observed; patient continued to be exposed to RHIs currently.

Motor signs: not observed.

Other psychiatric features: major depressive disorder and post-traumatic stress disorder, with symptoms of depression, anxiety, verbal irritability, and insomnia. Self-report measures suggested extremely severe depression, moderate anxiety, and stress.

Patient also suffers from severe migraine; Headache Impact Test (HIT-6) [34] score of 61.

*Provisional level of certainty for CTE Pathology:* possible CTE.

*Neuroimaging findings:* MRI data showed moderate PSPR volume loss and enlargement of the adjacent sulcus but without ventricular widening, hippocampal, or precuneal volume loss. PET findings indicated no evidence of hypometabolism in the bilateral parietal, striatal, frontal, and precuneal regions. However, there was mild hypometabolism in the lateral temporal region.

*Treatments and improvements:* over three years of treatment/consultation, patient's neuropsychiatric regimen was optimized to include mirtazapine, agomelatine, lamotrigine, and low dose lisdexamphetamine. Headache relief was achieved with calcitonin gene-related peptide antagonists and botulinum toxin type A. Donepezil led to subjective improvement, with her ACE-III score improving to 97/100.

***Case 3:***

*TES criteria I:* Patient is a former amateur boxer in his late 50s with a TER of 14 years from boxing and martial arts and 20 TRCs.

*TES criteria II:*

    Cognitive impairment: patient presented with a 3-year cognitive decline. ACE-III score of 72 with indications of impairments in processing, attention, registration, recall, calculation, paraphasic errors, and concrete proverb interpretation. Neuropsychological assessment showed average to low average intelligence with extremely severe memory impairment. He had significant difficulty in encoding new information, with scores on tests of new learning falling 3 standard deviations below

average. Recall after a 20 to 30-minute delay was at basal levels, and recognition memory was severely disordered. Executive dysfunction was also noted, particularly in cognitive flexibility and rapid word retrieval.

Neurobehavioral dysregulation: patient presented with a 10-year neuropsychiatric decline, characterized by anger responses to minimal triggers.

Progressive course: patient had 3 years of cognitive decline and 10 years of neuropsychiatric decline prior to first presentation.

*TES criteria III:* the patients' symptoms were not fully accounted for by other disorders.

*TES criteria IV:* mild dementia.

*Other supportive features:*

Delay onset: patient ceased exposure to RHI for 27 years prior to onset of symptoms.

Motor signs: Neurological examination revealed prominent bilateral paratonia and generalized akathisia. Impaired Luria sequencing was also noted.

Other psychiatric features: Self-report measures indicated moderate depression and mild stress, without elevated anxiety symptoms. DASS-21 score of 41 prior to Lamotrigine treatment.

*Provisional level of certainty for CTE Pathology:* probable CTE.

*Neuroimaging findings:* MRI data revealed mild PSP atrophy with ventricular enlargement and hippocampal volume loss, but no precuneal volume loss. PET findings showed moderate hypometabolism in the lateral temporal region, with parasagittal frontoparietal (PSFP) and PSPR changes.

*Treatments and improvements:* initial mood stabilisation with Lamotrigine, followed by

Donepezil upon mood improvements. ACE-III score improved to 83/100 on donepezil, consistent with marked subjective and objective improvement. DASS improved from 41 to 10 following Lamotrigine treatment.

*Case 4:*

*TES criteria I:* A male in his early 50s, with a TER of 18 years and 6 TRCs, including four motor vehicle accidents and one fall. Patient had been engaged in boxing since the age of 10.

*TES criteria II:*

Cognitive impairment: patient presented with short-term memory, insight, and recall impairment, and disorganisation. Neurological examination showed a paucity of language, with fragmentation and early paraphasic errors. ACE-III score of 94 with impaired phonemic verbal fluency, calculation, registration, and recall. Electroencephalography was unremarkable. Neuropsychological testing showed that patient's verbal and nonverbal intellect ranged from average to high average. However, patient exhibited significant difficulties in episodic memory, with new learning reduced and delayed recall falling up to 2.0 standard deviations below his intellect level. Executive function assessment revealed reduced problem-solving abilities, poor self-monitoring, and some perseverative tendencies. Patient's processing speed was mildly reduced.

Neurobehavioral dysregulation: patient presented with mood disturbances, irritability, anger to minimal triggers, hoarding, paranoia, and paraphilic tendencies unresponsive to SSRI treatment.

Progressive course: patient recalled experiencing continued cognitive decline for 10 years prior to presentation.

*TES criteria III*: the patients' symptoms were not fully accounted for by other disorders.

*TES criteria IV:* mild dementia.

*Supportive features for provisional levels of certainty for CTE pathology:*

Delay onset: patient ceased exposure to RHIs for 16 years prior to onset of symptoms.

Motor signs: Limb examination identified subtle right Parkinsonian features and mild axial rigidity.

Other psychiatric features: Self-report measures indicated extremely severe depression, extremely severe anxiety, and severe stress.

*Provisional level of certainty for CTE Pathology:* probable CTE.

*Neuroimaging findings:* MRI data revealed moderate PSPR volume loss and enlargement of the adjacent sulcus, with ventricular enlargement (AHI 0.39), hippocampal volume loss, but no precuneal volume loss. PET scans demonstrated bilateral parietal, striatal, frontal, precuneal and lateral temporal hypometabolism of mild severity.

*Treatments and improvements:* patient's neuropsychiatric regimen included sodium valproate and quetiapine, which were unhelpful, but lamotrigine was partially effective. Donepezil at 5 mg led to a good subjective response for memory improvement. A higher 10 mg daily dose of Donepezil exacerbated the patient's anger and impulsivity, raising safety concerns. and therefore, the dose was reduced to 5 mg ongoing.

**Case 5:**

*TES criteria I:* A male former rugby league player and professional boxer in his 40s with a TER of 26 years and over 100 TRCs.

*TES criteria II:*

Cognitive impairment: Patient presented with a 12-year cognitive decline, including impaired learning, insight, short-term memory, and language. Family history included frontotemporal dementia in the patient's mother and diagnosed Alzheimer's disease (AD) in the patient's father, who was also a boxer, and the clinical AD diagnosis might have been overlapping with possible CTE. ACE-III score of 88/100, showing difficulty in registration and recall, verbal fluency, and a tendency for perseveration. Neuropsychological assessment indicated marked cognitive impairment against a background of average intelligence. Episodic memory was severely disordered, with significant difficulty in encoding new material and recalling information after delays, but recognition memory was more preserved. The patient also showed marked difficulty in executive functioning, including strategic word retrieval, response inhibition, and sustained attention.

Neurobehavioral dysregulation: Patient presented with a 12-year neuropsychiatric decline in behaviour and mood, characterized by dysregulation and explosivity to minimal triggers, contrasted with self-reported prior personality. This change in behaviour caused significant stress in the patient's family. The patient also exhibited impulsivity and binge alcohol use correlated with the onset of neurocognitive decline, and sleep disturbance, but no RBD.

Progressive course: patient had 12 years of cognitive and neurobehavioral decline prior to first presentation.

*TES criteria III: not fully accounted for by other disorders*: the patients' symptoms were not fully accounted for by other disorders.

*TES criteria IV: level of functional dependence/dementia:* mild dementia

*Supportive features for provisional levels of certainty for CTE pathology:*

    Delay onset: not observed; patient continued to be exposed to RHIs to the onset of symptoms.

    Motor signs: Neurological examination revealed symmetrical parkinsonism with limb and axial involvement, mild rigidity with bradykinesia, and saccadic extraocular movements, but no tremor.

    Other psychiatric features: Self-report measures suggested moderate depression, with no significant anxiety or stress.

*Provisional level of certainty for CTE Pathology:* probable CTE.

*Neuroimaging findings:* MRI data revealed mild PSP atrophy without ventricular enlargement, or hippocampal or precuneal volume loss. PET findings indicated mild severity hypometabolism in the lateral temporal region and bilateral PSFP.

*Treatments and improvements:* Lamotrigine was commenced and up titrated to 100 mg twice daily which effectively stabilised the patient's mood. Donepezil treatment was associated with subjective improvements in cognition and mood stability.

### Case 6:

*TES criteria I:* A male in his 50s with a TER of 30 years through extreme sports participation, including snow sports and mountain biking, and 15 TRCs.

*TES criteria II:*

    Cognitive impairment: Patient presented with 10 years of cognitive decline prior to first presentation. Initial neuropsychological testing two years prior to presentation

showed impairment in verbal learning, memory, and psychomotor speed. Subsequent testing indicated impaired auditory attention, basic working memory, and cognitive flexibility. The most recent assessment (3 years after first presentation) revealed average to low average intellect with marked episodic memory impairment. The patient's ability to learn verbal material was significantly below average, with delayed recall and recognition memory consistent with immediate recall, indicating poor initial encoding. Learning and retention of visual memory were consistent with intellect. Executive function assessment showed difficulty in mental control and rapid word generation, with reduced processing speed. ACE-III score of 79.

    Neurobehavioral dysregulation: irritability, social isolation, and workplace dysfunction.

    Progressive course: Patient had 10 years of cognitive and neuropsychiatric decline prior to first presentation and current continued decline.

*TES criteria III:* the patients' symptoms were not fully accounted for by other disorders.

*TES criteria IV:* mild dementia.

*Supportive features for provisional levels of certainty for CTE pathology:*

    Delay onset: patient had continued RHI exposure through symptoms onset and current ongoing exposure.

    Motor signs: cranial nerve, limb, gait, and speech examinations were normal.

    Other psychiatric features: The patient exhibited abulia (loss of motivation) and depression. Self-report measures indicated severe depression, but no elevated anxiety or stress.

*Provisional level of certainty for CTE Pathology:* probable CTE.

*Neuroimaging findings:* MRI data revealed mild PSP atrophy without ventricular enlargement, hippocampal, or precuneal volume loss. PET findings indicated bilateral hypometabolism in the lateral temporal region, bi-lateral PSFP, and PSPR, all with mild severity.

*Treatments and improvements:* The patient was commenced on Donepezil for cognitive improvements and Lamotrigine for mood stabilisation, both of which the patient responded positively.

**Discussion**

The case series characterised TES's manifestation and the experience in symptomatic management in six TES patients in the Sydney area, Australia and suggested several antemortem neurological signatures of possible and probable CTEs. We identified two key radiological features in this case series: posterior superior parietal region (PSPR) volume loss and increased Evans index on clinical structural MRI (Figure 1) and specific patterns of hypometabolism on FDG-PET (Figure 2). This study did not use structural MRI or FDG-PET in a group comparison, voxel-wise, or region-of-interest-based quantitative analysis but instead relying solely on radiologist reports, highlighting the possibility of reliance on currently existing clinical capabilities to enhance TES diagnosis.

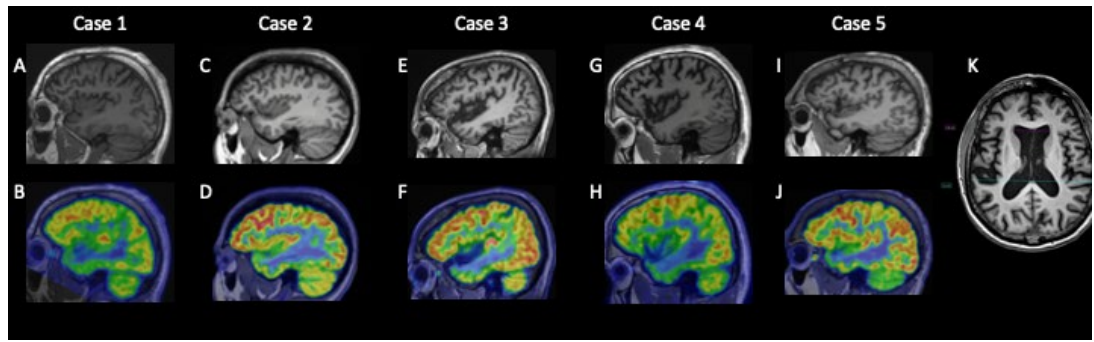

**Figure 1:** MRI and collocated FDG-PET findings showing Case 1-5 lateral temporal hypometabolism (A-J) and Posterior Superior Parietal volume loss with ventricular widening as measured by Evans Index (K).

While all the participants had possible or probable TES (5/6 probable TES), the radiologist-identified MRI features of ventricular widening (Evans Index > 0.3 or ventriculomegaly [35]) and hippocampal volume loss (hippocampal volume loss was a common feature in prior studies [12,14,19–21]) was not universal: only 2 probable CTE cases (case 3 and 4) had both; other four patients had none. Nevertheless, 5/6 cases had Evans Index > 0.25. None had precuneal volume loss, but all had PSPR atrophy and expansion of the adjacent sulcus, indicating PSPR atrophy being an early and/or distinctive feature of TES/CTE. PSPR atrophy may hold certain advantages over other features, for example, cavum septum pellucidum (CSP) [9]. While the presence of CSP was noted to be higher in TES [14–18], it is not specific to the disease: CSP has more evidence linking its presence to repetitive head trauma than the underlying CTE pathology [9]. CSP has also been found to be prevalent in mental disorders [36]; on the other hand, those with psychiatric conditions are also more likely to have head trauma [37,38].

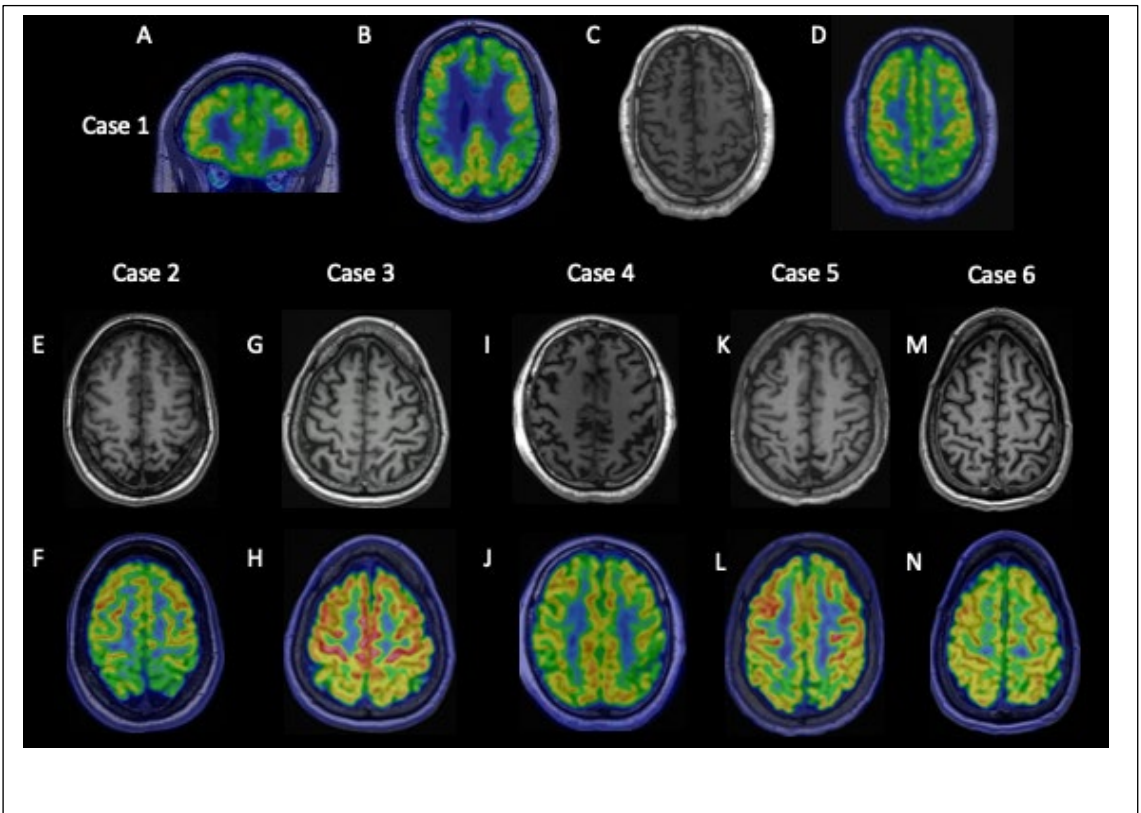

**Figure 2**: Representative case showing FDG-PET imaging of Anterior Frontal Parasagittal (AFPS) cortical hypometabolism (A) and (B) and Posterior Superior Parietal (PSP) volume loss in Cases 1-6 using MRI, correlating with FDG PET hypometabolism in Cases 1, 3, 4 and 5 (C-N).

Spatial patterns of neurologist-defined FDG-PET hypometabolism indicated that the most commonly involved areas included, in decreasing order, the anterior frontal areas (AF-AFPS, 5/6 cases and all 5 of the probable cases), the lateral temporal (LT, 5/6 cases and 4/5 of the probable cases), the posterior superior parietal region (PSPR, 4/6 cases and 4/5 of the probable cases). The medial temporal areas had hypometabolism less often (2/6 cases), though the cases where this area was involved tended to have more severe grades of the highest hypometabolism grading (1/2 moderate and 1/1 severe). The precuneal was not shown to have clinically detected hypometabolism. The hypometabolism pattern indicated that hypometabolism spread AF-AFPS → LT → PSPR which fit with the general progression of p-Tau as CTE progressed [2]. This progression underscores the disease's insidious nature and its potential overlap with other tauopathies like AD and some forms of Frontotemporal Lobar Degeneration. However, the absence of beta-amyloid neuritic plaques and the unique p-tau pathology distribution distinguish CTE from these conditions [39]; Flortaucipir ($^{18}$F) (FTP) can be used as differential diagnosis for AD vs. CTE [28].

Cases 3 and 4 who had MT FDG-PET hypometabolism and greatest highest severity grades of whole-brain hypometabolism also had clinically detected ventricular widening and Evans Index > 0.3. This suggested that they were further along the TES and probably CTE progression; more advanced CTE was associated with ventricular enlargement [9,40]. We are proposing the following TES and possible/probable CTE progression of clinically available and radiologist-interpreted imaging reports:

| | Radiological progression of TES/possible-probable CTE |
|---|---|
| 1 | Posterior Superior Parietal (PSPR) volume loss |
| 2 | Anterior Frontal Parasagittal - PSPR hypometabolism, 0.25 < Evans Index < 0.3 |
| 3 | Lateral progressing to medial temporal hypometabolism, Evans Index > 0.3 |

This study has limitations, first of which is the limited sample that mostly consisted of patients of one physician. While we propose a TES and possible/probable CTE progression of imaging features, the defining severity and progression of TES is still a work in progress and CTE can only be confirmed post-mortem. In our proposal, we used the assumption that TES features should follow an accumulating pattern across a cross-sectional sample of TES patients: the sample should contain patients at different condition stages, and later stage TES patients will present with similar features seen in earlier TES patients. Consequently, the most common features seen in the patient sample are more likely to be that of earlier stages while less common features are more likely to be that of later stages.

**Declarations**

*Ethics approval and consent to participate.*

Patients' clinical data, health records, and imaging reports are stored in the Clinic to Cloud practice management system (https://www.clinictocloud.com/). This case study uses a retrospective design and study ethics approval was obtained from the Macquarie University Human Research ethics committee (ethics approval number 6057) for data to be extracted from the database for analysis and publication conditioned upon patients' or substitute decision makers' informed consent for their deidentified data to be used. Written informed consent was obtained from all patients; after informed consent was granted, patients' data and reports were extracted from the database for the study.

The study was conducted in accordance with the *National Statement on Ethical Conduct in Human Research 2007 (Updated 2018)* [41].

*Data Availability Statement*

All data produced from this study, including extracted data tables and spreadsheets, will be made available upon appropriate request to the corresponding author.

*Competing interests*

The authors declare that they have no competing interests.

*Funding*

We would also like to acknowledge the Motor Accident Insurance Commission (MAIC) for supporting authors F.N and X.V.T (2014000857).

*Authors' contributions*

R.M is the neurologist treating the cases, collected the data, drafted the initial draft. F.N drafted the initial drafts and prepared the figures. X.V.T revised and edited the final version of the manuscript. J.M and E.H conducted the imaging and interpreted the imaging results. J.B and M.W collected the data. All authors reviewed the manuscript and approved the submission.

**Acknowledgements**

We would like to acknowledge Dr Margery Pardey, Chief Research Facilitator, Macquarie Medical Imaging for her role in data collection and research coordination.

updated-2018.

Table 1. Demographic and Trauma Exposure Features

TBI = Traumatic Brain Injury, TER = Total Exposure to Repetitive mild TBI, TRC = Total Recalled Concussions, MVA = Motor-Vehicle Accident, M = Male, F = Female.

| Case | Age | Gender | TBI source | TER (years) | TRC |
|---|---|---|---|---|---|
| 1 | Early 50s | M | Rugby Union, Rugby League, Boxing, Martial arts | 33 | 50+ |
| 2 | 30s | F | Self-injury, domestic violence | 26 | 100+ |
| 3 | Late 50s | M | Boxing | 14 | 20 |
| 4 | Early 50s | M | Boxing, Rugby Union, MVA | 18 | 6 |
| 5 | Early 40s | M | Rugby league, boxing | 26 | 100+ |
| 6 | Early 50s | M | Extreme sports | 30 | 15+ |

Table 2. Cognitive Features

More + = more deficits, - = no deficit observed

| Case | Cognitive decline duration (years) | ACE-III | Verbal Memory | Visual Memory | Verbal Fluency | Self-Monitoring | Processing Speed | Attention |
|---|---|---|---|---|---|---|---|---|
| 1 | 5 | 85 | ++ | + | ++ | + | + | + |
| 2 | 5 | 87 | - | ++ | - | - | + | ++ |
| 3 | 3 | 72 | +++ | - | + | + | + | + |
| 4 | 15 | 94 | +++ | - | + | + | + | + |
| 5 | 15 | 88 | +++ | - | ++ | ++ | - | + |
| 6 | 10 | 79 | ++ | - | ++ | ++ | +++ | + |

Table 3. Psychiatric Features

More + = more deficits, - = no deficit observed

| Case | Depression | Anxiety | Irritability | Behavioural Dysregulation | Paranoia | Psychotropic Agents |
|---|---|---|---|---|---|---|
| 1 | ++ | + | ++ | + | - | Fluvoxamine, Donepezil |
| 2 | ++ | + | ++ | ++ | - | Mirtazapine, Agomelatine, Lamotrigine, Lisdexamfetamine, Donepezil |
| 3 | ++ | + | +++ | + | - | Lamotrigine, Donepezil |
| 4 | ++ | ++ | +++ | ++ | ++ | Lamotrigine, Donepezil |
| 5 | + | - | ++ | ++ | - | Lamotrigine, Donepezil |
| 6 | ++ | - | + | - | - | Lamotrigine |

Table 4. MRI Features

PSP = Posterior Superior Parietal

| Case | PSP atrophy | PSP atrophy grading | Ventricular widening (Evans Index) | Hippocampal volume loss | Precuneal volume loss |
|---|---|---|---|---|---|
| 1 | Yes | Mild | No (0.27) | No | No |
| 2 | Yes | Mod | No (0.22) | No | No |
| 3 | Yes | Mild | Yes (0.30) | Yes | No |
| 4 | Yes | Mod | Yes (0.39) | Yes | No |
| 5 | Yes | Mild | No (0.28) | No | No |
| 6 | Yes | Mild | No (0.26) | No | No |

Table 5. Detected FDG-PET hypometabolism.

| Case | Bilateral AF | AFPS | Bilateral PSFP | PSPR | LT | MT | Higher hypometabolism severity in LT vs. MT | Precuneus | Highest hypometabolism grading in the brain |
|---|---|---|---|---|---|---|---|---|---|
| 1 | Yes | Yes | Yes | Yes | Yes | No | Yes | No | Mod |
| 2 | No | No | No | No | Yes | No | Yes | No | Mild |
| 3 | No | Yes | No | Yes | Yes | Yes | Yes | No | Moderate |
| 4 | Yes | Yes | Yes | Yes | Yes | Yes | Yes | No | Severe |
| 5 | Yes | Yes | Yes | Yes | Yes | No | Yes | No | Mild |
| 6 | Yes | Yes | No | No | No | No | No | No | Mild |

AF = Anterior Frontal, AFPS = Anterior Frontal Parasagittal, PSFP = Parasagittal Frontoparietal, PSPR = Posterior Superior Parietal Region, LT = Lateral Temporal, MT = Medial Temporal